%% file: XuRX.tex
\documentclass[12pt]{article}
\usepackage{amsmath}
\usepackage{amssymb}
\usepackage{txfonts}
\usepackage{epsfig}

\input econfmacros.tex
\def\Title#1{\begin{center} {\Large {\bf #1} } \end{center}}

\begin{document}

\Title{The state of cold quark matter:\\%
\vspace{2mm}%
{\em a model-independent view}}

\bigskip\bigskip


\begin{raggedright}

{\it Renxin Xu\index{Xu, R.}\\
Department of Astronomy\\
School of Physics\\
Peking University\\
Beijing 100871\\
P. R. China\\
{\tt Email: r.x.xu@pku.edu.cn}}
\bigskip\bigskip
\end{raggedright}

\begin{abstract}
From a model-independent point of view, we address the possibility
that quark clustering could occur in cold quark matter at realistic
baryon densities because of the likely strong coupling between
quarks in compact stars.
\end{abstract}

\section{Introduction}

In this paper, I would like to present an idea about the state of
cold quark matter, for your comments and suggestions. It is
generally supposed that color superconductivity (CSC) will occur in
dense quark matter. However, besides this CSC state, another one may
be possible: strong coupling could cause quarks to {\em cluster} in
realistic dense quark matter in which case a {\em solid} state of
quark matter would exist at low temperatures.

This conjecture may alleviate many problems that challenge
astronomers and astrophysicists. For instance, we still understand
neither the supernova explosion mechanisms, nor the central engines
of Gamma-ray bursts (GRBs). Also, what's the nature of pulsar-like
stars, especially the so-called magnetar candidates (AXPs, anomalous
X-ray pulsars, and SGRs, soft gamma-ray repeaters)?
The conjecture also has bearing on some related problems in physics.
For example, what are the properties of the strongly coupled
quark-gluon plasma (sQGP)? How can one draw the QCD (quantum
chromo-dynamics) phase diagram?
All of these problems, I think, may be related to a {\em big}
challenge: the physics of (cold) matter at supra-nuclear density.
In principle, that equation of state could be derived from the first
principles (i.e., QCD), however this is as yet not possible because
of the strong none-perturbative effects at low energy scale.
One requirement for solving the big problems is to have physicists
and astronomers to exchange information during their work when
trying to ``dig a tunnel''.

The outlines of this paper is as follows. In \S2, I will start with
an analogous but simple phase diagram, that of water (H$_2$O).
Then, in \S3, the conjectured QCD states of cold quark matter,
color-superconducting $v.s.$ quark-clustered, are discussed, and our
ideas to understand various observations are presented in \S4.
Finally, I summarize in \S5.

\section{An analogy: the H$_2$O phase diagram}

Let's begin with the phase diagram of water, rather than the QCD
phase diagram, since the H$_2$O phases make it easier for us to
understand the essential but similar physics.

Suppose I give you a large number of particles of electrons ($e$),
protons ($p$), and oxygen nuclei ($^{16}O$), at room temperature of
$T\sim 300$ K and density of $\rho\sim 1$ g/cm$^3$, with a number
proportion of
$$%
n_e:n_p:n_O = 10:2:1.
$$%
Certainly attractive and repulsive forces exist between those
particles. My question is: What's the state of such a particle
system?
You know the answer from experience: liquid water!
However, as a physicist who may have never seen water, can you
predict the state of this particle system by calculations, starting
from the laws of physics?

We can first compute the number densities of the particles,
$$%
n_e=3.35\times 10^{23}/{\rm cm}^3,\;\; n_p=n_e/5,\;\; n_O=n_e/10,
$$%
and also the distances between each kind of particle,
$$%
l_e=n_e^{-1/3}\sim 1\times 10^{-8}{\rm cm},\;\; l_p\sim 2\times
10^{-8}{\rm cm},\;\; l_O\sim 3\times 10^{-8}{\rm cm}.
$$%
Then we can also calculate their quantum wavelengths,
$$%
\lambda_e={h\over \sqrt{2m_ekT}}\sim 8\times 10^{-7}{\rm cm},\;\;
\lambda_p\sim 2\times 10^{-8}{\rm cm},\;\; \lambda_O\sim 4\times
10^{-9}{\rm cm}.
$$%
By comparing the distances and the wavelengths, you my conclude:
quantum Fermi-Dirac statistics applies to the electrons and protons,
while classical Maxwell-Boltzmann statistics applies to oxygen-16.

Furthermore, one can also estimate the chemical potentials of the
degenerate electrons and protons. Assuming, for simplicity, zero
temperature and ignoring the electromagnetic interactions, one comes
to
\begin{equation}
\left\{ \begin{aligned}
\mu_e&={\hbar^2\over 2m_e}(3\pi^2)^{2/3}\cdot n_e^{2/3}=18{\rm eV}=2\times 10^5{\rm K}\gg T,\\
\mu_p&={\hbar^2\over 2m_p}(3\pi^2)^{2/3}\cdot n_p^{2/3}=0.003{\rm
eV}=28{\rm K}\lesssim T.
\end{aligned}%
\right.%
\end{equation}
If interactions are neglected, the conclusions are then as follows:
\begin{equation}
\left\{ \begin{aligned}
Electrons: & \; \; \; {\rm strongly\; degenerate\; Fermi\; gas\; at\;} T \sim 0;\\
Protons: & \; \; \; {\rm weakly\; degenerate,\; classical\; gas};\\
O^{16}\; nuclei: & \; \; \; {\rm classical\; gas\; (M-B\;
statistics).}
\end{aligned}%
\right.%
\nonumber%
\end{equation}

However, these conclusions would be wrong, because interactions do
play an important role. Due to the electro-magnetic interaction,
particle clusters have to form in the system: 13 particles (10
electrons, two protons, and one oxygen nucleus) will clustered in a
group as one water molecular, H$_2$O.

Let's see how much further we could go using fundamental physics.
Considering simply a cluster composed of an electron ($e$) and a
proton ($p$), with a length scale $l_{\rm ep}$, from Heisenberg's
uncertainty relation, one infers a kinetic energy of $\sim
p^2/m_e\sim \hbar^2/(m_el_{\rm ep}^2)$, which has to be comparable
to the interaction energy of $E_{\rm ep}\sim e^2/l_{\rm ep}$ in
order to have a bound state. One then finds,
\begin{equation}
l_{\rm ep} \sim {\hbar^2/(m_ee^2)} = {1\over \alpha_{\rm em}} {\hbar
c\over m_ec^2},\; \; E_{\rm ep} \sim \alpha_{\rm em}{\hbar c\over
l_{\rm ep}} = \alpha_{\rm em}^2m_ec^2,
\label{EM_cluster}%
\end{equation}
where $\alpha_{\rm em}=e^2/(\hbar c)=1/137$ is the coupling constant
of the electro-magnetic interaction. Such a simple estimate is very
effective because we have $l_{\rm ep}=0.53~\AA$, which is very close
to the Bohr radius $a_{\rm B}=0.52~\AA$, and interaction energy
$E_{\rm ep}=27$ eV, approximately equals to the Rydberg constant
$R_{\rm H}=13.6$ eV. Therefore, one has to modify the previous
conclusion: there will be clusters of Hydrogen atoms ($e$+$p$) [as
well as Oxygen atoms (8$e$+$^{16}O$)] in the particle system because
$E_{\rm ep}\gtrsim {\rm Max}[\mu_e, \mu_p]$.

What's next? What's the real phase diagram of the system? This is
still a challenging problem to solve from first principles only, and
we have to use experiential knowledge.
Certainly, we know that larger clusters (H$_2$O, i.e., the water
molecules) will form if the residual electromagnetic interaction
between H- and O-atoms is considered. Note that we are still {\em
far away} from our final goal of reproducing the global phase
diagram calculated from first principles, even though we know the
elementary composition of the system, H$_2$O.
Fortunately, we have this diagram although it is a long and hard
work for experimentalists to obtain.

From above, we see that it is very difficult to infer the properties
even for the very simple case of water.
Actually, for any electromagnetic particle system, a general
calculation of matter properties, including the phase diagram, is
too complex to be possible because of the interaction, even though
it has coupling $\alpha_{\rm em}\ll 1$.
Indeed, fascinating new quantum phenomena are continually emerging,
which are related to the degenerate quantum gas, and a recent
conference was focused on these topics
(http://coldatom.castu.tsinghua.edu.cn/).

And then, what about the QCD phase diagram for quarks, for which the
coupling parameters may well be close to and possibly even greater
than $1$?

\section{Cold quark matter: color-superconducting or quark-clustering?}

In this section, I would like to present a few points about the
states of QCD matter, which I hope will lead to comments and
suggestions.

Certainly, due to the nature of asymptotic freedom, when temperature
$T$ or baryon chemical potential $\mu_{\rm B}$ are high in the QCD
phase diagram, there is a phase in which quarks are deconfined. At
the high density, low temperature regime on which we are focused,
cold dense quark matter could be of Fermi gas or liquid {\em if} the
interaction between quarks is negligible. However, the questions is:
can the density in realistic compact stars be so high that we can
neglect the interaction?

The average density of a pulsar-like star with a mass of 1.4
$M_\odot$ and a radius of 10 km is $2.4\rho_0$, where $\rho_0$ is
the nuclear density.
Even at this density, the quark degree of freedom could appear since
the critical density $n_{\rm c}$ to break nucleons,
$$%
n_{\rm c}={3\over 4\pi r_n^3}\simeq 1.5r_1^{-3}\rho_0,
$$%
is only about $2\rho_0$ if $r_1\equiv r_n/(1~{\rm fm})=0.9$, where
$r_n$ denotes the nucleon radius. It could then be reasonable to
consider realistic dense matter in compact stars already as quark
matter.

For three-flavor quark matter with density of, e.g., $\rho=3\rho_0$,
one may make estimates similar to those done above for water.
We have number densities for each flavor of quark, $u$, $d$, and
$s$, of $n_u\simeq n_d\simeq n_s\sim (3\times 0.16 = 0.48)$
fm$^{-3}\sim 5\times 10^{38}$/cm$^3$, and typical separations,
$$%
l_u\simeq l_d\simeq l_s\sim 1.3\; {\rm fm}.
$$%
The typical distance between quarks is then $l_Q\simeq l_u/(3^{1/3})
\sim 0.9$ fm. The quantum wave-length, however, is order of
$$%
\lambda_q\simeq {h\over \sqrt{2m_qkT}}=5\times
10^3m_{300}^{-1/2}T_6^{-1/2}\; {\rm fm}\gg l_Q
$$%
where quarks are assumed to be dressed, with mass of
$m_q=m_{300}\times 300$ MeV, and $T_6=T/(10^6\; {\rm K})$.
So quantum Fermi-Dirac statistics seems to apply to the case of cold
quark matter. By turning off the interaction, a further calculation
of
the quark chemical potential at zero temperature shows %
\begin{equation}
\mu_u^{\rm NR}\simeq \mu_d^{\rm NR}\sim \mu_s^{\rm NR}\approx {\hbar^2\over 2m_q}(3\pi^2)^{2/3}\cdot n_u^{2/3}=380{\rm MeV}\gg T\\
\label{mua}%
\end{equation}
if quarks are considered moving non-relativistically, or
\begin{equation}
\mu_u^{\rm ER}\simeq \hbar c (3\pi^2)^{1/3}\cdot n_u^{1/3}=480{\rm MeV}\gg T\\
\label{mub}%
\end{equation}
if quarks are considered moving extremely relativistically.
Thus, with this estimate of a quark chemical potential order of 0.4
GeV, much higher than thermal kinetic energy, we would infer that
realistic dense quark matter is a strongly degenerate Fermi gas.

However, as was shown in the last section for the case of water,
interactions between quarks near the Fermi surface may play an
important role. Similar to the condensed matter physics of
low-temperature superconductivity, quark pairs may form due to
attraction mediated by gluons, and cold quark matter may in a
BCS-like color superconductivity (CSC) state (see~\cite{csc} for a
recent review).
Such a CSC solution was found in perturbative quantum
chromo-dynamics (pQCD) at extremely high density, and is supposed to
exist in realistic quark matter of compact stars according QCD-based
effective models.

But, what if the color-interaction is strong rather than weak?
Similar to the case of clustered water molecular discussed
previously, would the strong interaction result in the formation of
quark clusters?
There are repulsive and attractive interactions between two quarks,
which is very analogous to the $\{e,p,O\}$-system presented in \S2.
A similar estimate to Eq.(\ref{EM_cluster}) for the length scale
($l_q$) and interaction energy ($E_q$) in a quark cluster gives if
quarks are dressed,
\begin{equation}
l_q \sim {1\over \alpha_s} {\hbar c\over m_qc^2}\simeq {1\over
\alpha_s}\; {\rm fm},\; \; E_q \sim \alpha_s^2m_qc^2\simeq
300\alpha_s^2\; {\rm MeV}.
\label{color_cluster}%
\end{equation}
This is dangerous for the CSC state since $E_q$ is approaching and
even greater than $\sim 400$ MeV if $\alpha_s>1$. So, what do we
know about $\alpha_s$ for realistic dense quark matter from recent
work on perturbative and non-perturbative QCD?

At extremely high density, the QCD coupling decreases as density
increases, approximately as~\cite{running}
\begin{equation}
\alpha_{\rm s}(\mu)\equiv {g_{\rm s}^2\over 4\pi} \approx{1\over
\beta_0\ln(\mu^2/\Lambda^2)}%
\label{alpha}
\end{equation}
in pQCD, where $\beta_0=(11-2n_f/3)/(4\pi)$, $n_f$ is the number of
quark flavors, the renormalization parameter $\Lambda=(200\sim 300)$
MeV, and the energy scale $\mu$ likely of order the chemical
potential estimated in Eq.(\ref{mua}) and Eq.(\ref{mub}).
One sees from Eq.(\ref{alpha}) that $\alpha_s\sim 0.1\gg \alpha_{\rm
em}=1/137$ even if the density is unreasonably high, $\sim
10^6\rho_0$. However, Eq.(\ref{alpha}) cannot applicable when $\rho
< 10\rho_0$, since there non-perturbative effects dominate.

How about $\alpha_s$ in non-perturbative QCD? This is a challenging
problem. Indeed, non-perturbative QCD is related to one of the seven
Millennium Prize Problems named by the Clay Mathematical Institute.
Certainly, the coupling is strong at low energies, and the
perturbative formulation is not applicable. Nevertheless, there are
some approaches to the non-perturbative effects of QCD, one of which
uses the Dyson-Schwinger equations (DSE).
This method was tried by Fischer et al.~\cite{dse1,dse2}, who
formed,
\begin{equation}
\alpha_s(x)={\alpha_s(0)\over \ln(e+a_1x^{a_2}+b_1x^{b_2})},
\label{dse}%
\end{equation}
where $a_1=5.292~{\rm GeV}^{-2a_2}$, $a_2=2.324$, $b_1=0.034~{\rm
GeV}^{-2b_2}$, $b_2=3.169$, $x=p^2$ with $p$ the typical momentum in
GeV, and that $\alpha_s$ freezes at $\alpha_s(0)=2.972$.
For our case of dense quark matter at $\sim 3\rho_0$, the chemical
potential is $\sim 0.4$ GeV [Eqs.(\ref{mua}) and (\ref{mub})], and
thus $p^2\simeq 0.16$ GeV$^2$. Thus, it appears that the coupling in
realistic dense quark matter should be greater than 2, being close
to 3 in the Fischer's estimate presented in Eq.(\ref{dse}).
Therefore, $E_q>\mu_u$, and quarks should be clustered in realistic
dense quark matter, as was found in the case for water in \S2.

What kind of quark clusters could exist in cold quark matter?
Generally speaking, as quarks are Fermions, the exchange of two
quarks in position and inner spaces should change the sign of the
wave function of the quantum system of the particles. For a quark
system, the interaction via attraction channel may result in the
formation of quark clusters if the wave function $\Psi(q_1,q_2,...)$
is symmetric when exchanging two quarks in position space, but
$\Psi(q_1,q_2,...)$ should be asymmetric in inner (e.g., spin,
flavor, and color) spaces.
The $Q_\alpha$-cluster~\cite{xu03} has such a wave function, in
which the spin states ($\{ |1/2 \rangle,\; |-1/2 \rangle$) are 2,
the flavor states ($\{u,\; d,\; s\}$) are 3, and the color states
($\{r,\; g,\; b\}$) are 3 in one ground state of position space. The
total number of quarks in a $Q_\alpha$-cluster is then $2\times
3\times 3=18$.
Certainly di-$Q_\alpha$-clusters ($Q_{2 \alpha}$),
tri-$Q_\alpha$-clusters ($Q_{3 \alpha}$), and even more massive
clusters could also be possible due to the attraction between quarks
or $Q_\alpha$-clusters.

In dense quark matter at $\rho \sim 3\rho_0$, the distance between
$Q_\alpha$-clusters is about
$$%
l_{Q_\alpha}\sim ({3\times 0.48\over 18})^{-1/3}\simeq 2~{\rm fm},
$$%
while the cluster length would be $l_q\sim 1$fm/$\alpha_s$ [from
Eq.(\ref{color_cluster})], which could be smaller than
$l_{Q_\alpha}$. However, now the question arises about what state
these clusters would be in: could the quark-cluster's quantum
wave-length be much longer than $l_{Q_\alpha}$? The answer is
``yes'' if the interaction between the clusters is negligible.
But there will be residual interaction between the clusters, and we
need such an interaction to have associated energies of
$$%
E_{\rm cluster} \sim {\hbar^2 \over 2m_{Q_\alpha}l_{Q_\alpha}^2}
\sim 1~{\rm MeV}
$$%
to localize the clusters (where the mass of non-relativistic
clusters $m_{Q_\alpha}\simeq 300\times 18$ MeV). Thus, we infer that
the potential drop between clusters has to be deeper than $\sim 1$
MeV in order to have classical quark-clusters, rather than a cluster
quantum gas. This could be reliable since the energy scale of strong
interaction is generally higher than 1 MeV (e.g., in nuclear
physics).

How and where can we experimentally test and determine the QCD
phases of cold quark matter? Observations of pulsar-like stars are
certainly useful, and in \S4 one will see that we could need a solid
state of quark stars to help us understand different observed
manifestations.

\section{To understand different manifestations of pulsar-like stars in quark star models}

What can quark-clustering do for us? I think there are at least four
advantages, which I will discussed below.

{\em A stiffer equation of state.}
It is conventionally though that the maximum mass of quark stars
should be lower than that of neutron stars, because the quarks in
quark matter (e.g., in the MIT bag model) are relativistic and then
the equation of state of quark matter is soft. However, quark
clusters move non-relativistically in our case, and hence the
equation of state of clustering quark matter is stiff, likely even
stiffer than that of hadronic matter. It may well be possible to
obtain a maximum mass of $\geq 2M_\odot$. Normal neutron star models
could be ruled out but a quark star model would be preferred if
astronomers detect a pulsar-like star with mass higher than the
maximum mass of reliable neutron star models.

{\em Solidified quark matter.}
\begin{figure}
  \centering
    \includegraphics[width=10cm]{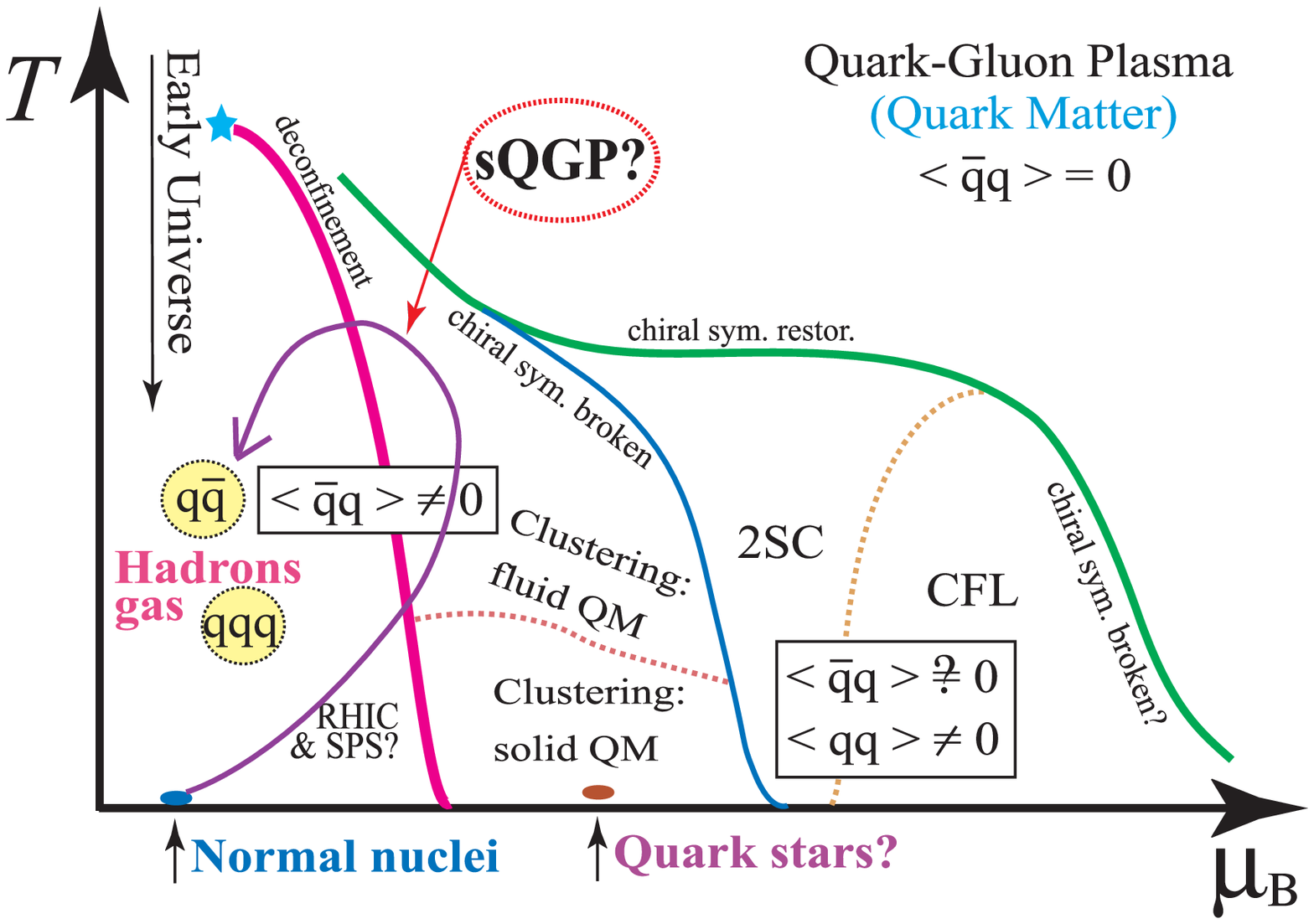}
    \caption{%
Schematic illustration of the QCD phase diagram speculated from
different manifestations of astrophysical compact stars. A quark
cluster state is conjectured when chiral symmetry is broken though
quarks are still unconfined. The cold quark matter with quark
clusters should be in a solid state at low temperature.
\label{QCDphase}}
\end{figure}
Similar to normal matter, dense quark matter could also be in a
solid state if the interaction energy between quark clusters is much
higher than the kinetic energy, $kT$. Two kinds of solid would be
possible, dependent on the penetration probability of quark clusters
through barriers between them. A {\it classical} solid may form if
the barrier penetration is negligible in case of strong interaction,
while a {\it quantum} solid could exist if the penetration is
significant in case of weak interaction.
An astrophysically conjectured QCD phase diagram, with the inclusion
of such a solid state of quark matter, is shown in
Fig.~\ref{QCDphase}. As the baryon density decreases, the states of
cold quark matter may change from BCS to BEC and even
quark-clustering phases.
A solid state of dense quark matter could help understand pulsar
glitch, precession, and even a Planckian thermal spectrum. The idea
of quark clustering (quark-molecular) could also be tested using the
strongly coupled quark gluon plasmas (sQGP) created in relativistic
heavy ion colliders. Certainly quark clusters are in a liquid state
because of high temperature of collider experiments.

{\em Energy release during a starquake.} Elastic and gravitational
energies could develop when solid stars evolve, and would be
released since a solid stellar object would inevitably have
starquakes when the strain energy increases to a critical value.
Assuming the two kinds of energies are of a same order, one could
have a huge energy release, of
$$%
\Delta E\simeq {GM^2\over R}\cdot {\Delta R\over R}\sim 10^{53}
{\Delta R\over R}~~~{\rm erg},
$$%
during star-quakes if the stellar radius changes from $R$ to $\sim
(R-\Delta R)$ (and momentum of inertia changes accordingly from $I$
to $\sim I(1-2\Delta R/R)$). Such kind of starquakes may be
responsible for the bursts and glitches observed in soft gamma-ray
repeaters (SGRs) and anomalous X-ray pulsars (AXPs).

{\em Ferro-magnetism?} What is the origin of the strong magnetic
fields of pulsars? This is still a matter of debate more than 40
years after the discovery.
In solid quark stars, ferro-magnetism might be the origin of the
magnetic field, and such a strong field would not decay
significantly (as observed in radio pulsars). It would seem very
worthwhile to study the spontaneously breaking mechanism of magnetic
symmetry in solid quark matter.

\vspace{2mm}

We have already done some modeling for different manifestations of
pulsar-like stars in solid quark star regime, and compared neutron
and quark star models. We find that the quark star model is
attractive. For instance, from the drifting subpulse phenomenon of
radio pulsars, one infers that the particles on a pulsar's surface
are strongly bound in the standard Ruderman-Sutherland model, with
binding energy $E_b\gtrsim 10$ keV, but the expected energy is
$E_b\lesssim 1$ keV for normal neutron stars or crusted strange
quark stars. However, the binding energy would be effectively
infinite for both positively ($u$-quarks) and negatively ($d$- and
$s$-quarks, and electrons) charged particles on the quark surface,
and we~\cite{xqz99} thus suggested that radio pulsars should be bare
strange quark stars in order to solve completely the ``binding
energy problem'' first posed ten years ago.

\section{Conclusions}

We suggest that realistic cold quark matter in compact stars would
well be in a solid state, where quark clustering occurs because of
strong coupling between quarks of quark matter at only a few times
of nuclear density.
At the same time, a solid quark-star model could help us understand
different manifestations observed in pulsar-like stars.
In the future, in order to fully and globally know the real QCD
phases, it is essential to combine three approaches: lattice QCD,
QCD-based effective models (e.g. DSE, NJL), and phenomenological
models (e.g. in astrophysics).
A solid state of cold quark matter may well have physical
implications in the research of fundamental strong interaction.

\bigskip
I would like to thank Prof. Marten van Kerkwijk very much for his
substantial help to improve the language as well as his valuable
scientific discussions during his visiting KIAA, and to acknowledge
useful discussions at our pulsar group of PKU.
This work is supported by NSFC (10778611, 10973002), the National
Basic Research Program of China (grant 2009CB824800) and LCWR
(LHXZ200602).

\def\Discussion{
\setlength{\parskip}{0.3cm}\setlength{\parindent}{0.0cm}
     \bigskip\bigskip      {\Large {\bf Discussion}} \bigskip}
\def\speaker#1{{\bf #1:}\ }
\def\endDiscussion{}

\Discussion

\speaker{J. Wambach} I am not quite sure if necessarily strong
interactions mean that you have clustering.  When you think of
nuclear matter, where the interaction between two nucleons is very
strong, so that from perturbation theory we cannot get anything. But
nuclear matter, as far as we know, is a liquid, and we do not need
clusters when the density becomes very low.  So somehow it is not
clear to me that it is necessary that if you have very strong
interactions, you automatically get clusters.

\speaker{R. Xu} Nuclear matter at the nuclear saturation density
should be in a liquid state because nuclei can be well understood in
the nuclear liquid model. Nevertheless, the degree of freedom of
nuclear matter at low density is nucleon, while the degree of quark
matter at higher densities is quark and gluon. The interaction of
the quark matter might then be stronger than that of nuclear matter.
Even in some heavy nuclei, $\alpha$-clusters are supposed to exist,
and an $\alpha$-decay would occur if an $\alpha$-cluster penetrates
the Coulomb barrier.
Additionally, a quark molecular model is suggested to explain the
features of sQGP detected in RHIC experiment. My point is: if quarks
could be clustered in dense and cold quark matter, it should be very
natural for us to expect to form quark molecular in the experimental
hot quark matter at higher temperature.

\speaker{T. Schaefer} I had sort of a similar comment about making
it a solid.  In principle it should be much easier to make nuclear
matter into a solid than quark matter, because the conditions are
much more favourable, the particles are much heavier and you have
repulsive short-distance interactions.  But not even nuclear matter,
under standard conditions, ever solidifies. So it seems to me that,
the higher you go in density, the harder it should be to make it a
solid.  Because quarks are mass-less, it takes a lot of kinetic
motion.  So, firstly, it is not clear how you would solidify this.

\speaker{R. Xu} So you believe the nuclear matter to be in a solid
state?

\speaker{T. Schaefer} No, we know that we cannot make nucleons in a
solid, so for nucleons it already doesn't work, and for quarks it
should be harder still.

\speaker{J. Horvath} Let me add a comment on that.  He doesn't have
actually the quarks, themselves, but he has a cluster in a single
object that has a mass of several times the mass of a nucleon.
Therefore, what is clustered is the quarks and the basic object is
now a cluster.  So, the clusters are in a periodic structure, if I
understood it correctly.  So the nucleons already don't show a
solid, this does not argue that the quarks cannot. The quarks are
inside the clusters, so he takes the clusters as a whole. Correct?

\speaker{R. Xu} Yes. The degrees would change from nucleon to quark
as the density increases. Only residual color interaction exists
between nucleons, while color-charged ones between quarks. The very
strong coupling between quarks may results in the formation of
quark-clusters as a new degree of freedom at a few nuclear
densities.

\speaker{R. Ouyed} The density you showed is three times saturation
density, which is very close to the average density of any compact
star, so in your model then there is no such thing as a neutron star
(or little room for neutron stars, if this phase exists).

\speaker{R. Xu} Yes, our picture is very simple, only quark stars
exist, and we are trying to understand different manifestations of
pulsar-like stars in this regime these years, although it is still
very difficult to calculate the critical density of de-confinement
phase transition because of non-perturbative coupling there.

\speaker{V. de la Incera} My question is what the role of the gluons
is here.

\speaker{R. Xu} Although the gluon degree can not be negligible in
hot quark matter, it could be integrated as the interaction between
quarks in our cold quark matter. So the gluon themselves may only
play an important role at high temperature.

\speaker{T. Schaefer} Let me show another sort of tiny comment.  The
cluster you wrote down, this eighteen quark state, in terms of
quantum numbers is what people used to call H dibaryons, so in some
sense what you suggest is H dibaryon fluid.  And that indeed seems
incredibly reasonable, a great idea, except that people could never
really find any evidence for this particular cluster that seemed so
incredibly favourable in terms of its quantum numbers.

\speaker{R. Xu} Yes, the study of multi-quark particles has a long
history, and it is recently a hot topic to search the experimental
evidence of such new hadron states (e.g., the pentaquark). Thought
multi-quark particles could be similar to the quark clusters
discussed here, they are in different kinds of vacuum and the
quark-clusters may be populated in quark matter but is very
difficult to exist as hadrons because of decaying to other lighter
hadrons in our vacuum.
One point could be relevant to electron's participating in our
vacuum, so that massive quark clusters are unstable in our daily
life but would be stable in cold strange quark matter where
electricity is negligible.
Certainly, both H-dibaryon-like and $Q_\alpha$-like quark clusters
are candidate degrees of freedom in cold quark matter.

\speaker{B. Zhang} A question from an astrophysical point of view.
In terms of sub-pulse drifting pulsars you talked about, the main
astrophysics evidence in favour of neutron stars these days is the
small hot spot.  The old problem of the quark stars was that it is
hard to concentrate the heat in a small cap, so do you have a new
idea about how to solve the problem?

\speaker{R. Xu} Your question is related to the thermal conductivity
of cold quark matter, which is still not certain up to now.
The thermal conductivity of liquid-like or gas-like cold quark
matter is high, but it could be low for solid quark matter because
of quarks being clustered and possible electron-phonon interaction.
Solid quark stars would have hot spots if the thermal conductivity
is not as high as speculated in conventional literatures.

\speaker{P. Zhuang} I think probably we can ask people to calculate
the probability of a pentaquark at high density. If this probability
increases with increasing density, then probably this lends support
to your idea. We can ask, for instance, Boqiang to calculate the
pentaquark at three nuclear density, and to see what happens.

\speaker{Y. Liu} A group in Nankai University had performed such
types of calculations. The results showed that the probability
decreased quite rapidly.

\speaker{P. Zhuang} The question is: if you try to consider the
lattice QCD at high density, this is a big problem and is not easy
to solve.

\endDiscussion
\end{document}

%% file: econfmacros.tex
\def\beq{\begin{equation}}
\def\eeq#1{\label{#1}\end{equation}}
\def\eeqn{\end{equation}}

\def\beqa{\begin{eqnarray}}
\def\eeqa#1{\label{#1}\end{eqnarray}}
\def\eeqan{\end{eqnarray}}

\let\bar=\overbar

\def\Dslash{\not{\hbox{\kern-4pt $D$}}}
\def\dslash{\not{\hbox{\kern-2pt $\del$}}}

\def\msb{{\bar{\ssstyle M \kern -1pt S}}}

\usepackage{fancyhdr,graphicx}

%% file: XuRX.bbl
\begin{thebibliography}{99}


\bibitem{csc}
M. G. Alford, K. Rajagopal, T. Schaefer, A. Schmitt, Rev. Mod.
Phys., {\bf 80}, 1455 (2008).

\bibitem{running}
G. M. Prosperi, M. Raciti, C. Simolo, Prog. Part. \& Nucl. Phys.,
{\bf 58} 387 (2007).

\bibitem{dse1}
C. S. Fischer, R. Alkofer, Phys. Lett., {\bf B536} 177 (2002).

\bibitem{dse2}
C. S. Fischer, J. Phys. G: Part. Nucl. Phys, {\bf 32} R253 (2006).

\bibitem{xu03}
R. X. Xu, ApJ, {\bf 596} L59 (2003).

\bibitem{xqz99}
R. X. Xu, G. J. Qiao, B. Zhang, ApJ, {\bf 522} L109 (1999).

\end{thebibliography}
